# (Curvature)$^2$-Terms for Supergravity in Three Dimensions

Hitoshi NISHINO[1]  and  Subhash RAJPOOT[2]

*Department of Physics & Astronomy*
*California State University*
*1250 Bellflower Boulevard*
*Long Beach, CA 90840*

**Abstract**

We investigate the effect of (Curvature)$^2$-terms on $N=1$ and $N=2$ supergravity in three dimensions. We use the off-shell component fields $(e_\mu{}^m, \psi_\mu, S)$ for $N=1$ and $(e_\mu{}^m, \psi_\mu, \psi_\mu^*, A_\mu, B, B^*)$ for $N=2$ supergravity. The $S$, $A_\mu$ and $B$ are respectively a real scalar, a real vector and a complex scalar auxiliary fields. Both for $N=1$ and $N=2$, only two invariant actions for (Curvature)$^2$-terms exist, while only the actions with (Scalar Curvature)$^2$ are free of negative energy ghosts. Interestingly, the originally non-physical graviton and gravitino fields start propagating, together with the scalar field $S$ for the $N=1$ case, or the complex scalar $B$ and the longitudinal component $\partial_\mu A^\mu$ for $N=2$. These new propagating fields form two new physical massive supermultiplets of spins $\left(\frac{1}{2}, 0\right)$ with $2\times(1+1)$ degrees of freedom for the $N=1$ case, and two physical massive $N=2$ supermultiplets of spins $\left(\frac{1}{2}, \frac{1}{2}, 0, 0\right)$ with $2\times(2+2)$ degrees of freedom for the $N=2$ case.

PACS: 04.65.+e, 11.30.Pb, 04.60.Kz

Key Words: Supergravity, Extended Supergravity, Three Dimensions, (Curvature)$^2$-Terms,
Quantizations, Graviton, Gravitino, Supersymmetry.

[1] E-Mail: hnishino@csulb.edu
[2] E-Mail: rajpoot@csulb.edu

## 1. Introduction

It has been well-known that the graviton in three dimensions (3D) is not physical, or has no actual degree of freedom [1]. Another way of expressing it is that the physical phase space of 3D gravity is related to the moduli space of flat $SL(2, \mathbb{R}) \approx SO(2,1)$ bundles [2]. It is also supported from the non-independence of the Riemann tensor $R_{\mu\nu\rho\sigma}$ from the Ricci tensor $R_{\mu\nu}$ and scalar curvature $R$, so that the field equation $R_{\mu\nu} = 0$ implies that $R_{\mu\nu\rho\sigma} = 0$.

If this is the case, then the question of the effect of (Curvature)$^2$-terms [3] on 3D gravity seems trivial, because all the possible (Curvature)$^2$-terms are either $(R_{\mu\nu})^2$ or $R^2$, both of which vanish on-shell upon the gravitational field equation $R_{\mu\nu} = 0$.

However, as a simple analysis reveals, there is a flaw in this argument. This is because even though the original graviton and gravitino do not propagate, such new additional (Curvature)$^2$-terms with higher-derivatives make them propagate, resulting in a completely different physical spectrum.

In this paper, we address ourselves to this subtle issue with curvature-square terms in supergravity in 3D. We first fix all the possible supersymmetric extensions of (Curvature)$^2$-terms for $N = 1$ [4] and $N = 2$ supergravity [5] [6] in 3D based on the off-shell multiplet $(e_\mu{}^m, \psi_\mu, S)$ and $(e_\mu{}^m, \psi_\mu, \psi_\mu^*, A_\mu, B, B^*)$, respectively. The latter has essentially the same auxiliary fields as $N = 1$ supergravity in 4D [7][8][9][10].

We next analyze the spin contents of each term by projection operators. Subsequently, we investigate the possible propagators with the right spin content and physical components under supersymmetry.

Interestingly, we will see that out of two possible supersymmetrizations of (Curvature)$^2$-terms, only that of the (Scalar Curvature)$^2$-term will have no negative energy ghosts, as desired both for $N = 1$ and $N = 2$. We will also see both for $N = 1$ and $N = 2$ that the originally frozen components of the graviton and gravitino start propagating and form massive supermultiplets consistent with supersymmetry.

## 2. Graviton and Gravitino in 3D

We start with the usual concept of graviton and gravitino in 3D. The common wisdom is that graviton and gravitino have no physical degree of freedom in 3D. In other words, there is no propagation of graviton or gravitino in 3D. One way of seeing this, *e.g.*, is to analyze polarization-tensor [11] based on Newman-Penrose formalism [12]. We provide below four different viewpoints (i) - (iv) to understand this fact, even though some of them have been already known as common wisdom.

(i) The first method is the simplest, *i.e.*, to count their on-shell degrees of freedom, as $(3-2) \times (3-1)/2 - 1 = 0$ for the graviton, and $(3-3) \times 1 = 0$ for the gravitino. The factor $(3-2)$ for the graviton is due to the deletion of the longitudinal and 0-th component, while the multiplication by $(3-1)$ with the division by 2 is for the symmetry. The final subtraction by 1 is due to the tracelessness. The factor $(3-3)$ for the gravitino is for the longitudinal component $(3-2)$ together with the $\gamma$-traceless condition.

(ii) The second method is to consider the fact that the Riemann tensor $R_{\mu\nu}{}^{\rho\sigma}$ in 3D is no longer independent of the Ricci tensor and the scalar curvature $R$, related by

$$R_{\mu\nu}{}^{\rho\sigma} \equiv +4\delta_{[\mu}{}^{[\rho}R_{\nu]}{}^{\sigma]} - \delta_{[\mu}{}^{\rho}\delta_{\nu]}{}^{\sigma}R \ . \tag{2.1}$$

Therefore, the 'on-shell' vanishing Einstein tensor $R_{\mu\nu} - (1/2)g_{\mu\nu}R \doteq 0$[3] implies the vanishing of the Ricci tensor $R_{\mu\nu} \doteq 0$, and therefore that of the Riemann tensor itself. Once the Riemann tensor vanishes on-shell, there is no degree of freedom left for the dreibein.

(iii) The third method is based on 'Coulomb gauge' analysis for the linearized gravity. To this end, we review the usual Coulomb gauge condition for an $U(1)$ gauge field in 3D:[4]

$$\partial_i A^i \stackrel{*}{=} 0 \qquad (i = 1, 2) \ . \tag{2.2}$$

Using this in the gauge-invariant field equation[5]

$$\Box A_\mu - \partial_\mu \partial_\nu A^\nu \doteq 0 \ , \tag{2.3}$$

we get

$$\partial_i^2 A_0 \doteq 0 \ . \tag{2.4}$$

If there is no singularity anywhere in the 2D subspace, there is no other solution than the trivial one $A_0 \doteq 0$, according to the Gauss's law, under the boundary condition $A_0 \to 0$ at spatial infinity.

Analogously to this Coulomb gauge condition on $A_\mu$, we can impose the condition on the linearized gravitational field $h_{\mu m} \equiv e_{\mu m} - \eta_{\mu m}$

$$\partial_i h_i{}^\mu \stackrel{*}{=} 0 \ . \tag{2.5}$$

together with the tracelessness condition

$$h \equiv h_\mu{}^\mu \stackrel{*}{=} 0 \ . \tag{2.6}$$

---

[3] We use the symbol $\doteq$ for a field equation or a solution, but not an algebraic identity.

[4] The indices $i, j, \cdots = 1, 2$ are for spatial 2D subspace. We use the symbol $\stackrel{*}{=}$ for a subsidiary condition.

[5] Our space-time metric is $(\eta_{\mu\nu}) = \text{diag.}(-,+,+)$. We use the symbol $\Box \equiv \partial_\mu^2 \equiv \eta^{\mu\nu}\partial_\mu\partial_\nu$ even in 3D.

Now the Einstein gravitational field equation at the linear order, which is equivalent to the vanishing Ricci tensor equation

$$R_{\mu\nu}\big|_{\text{Linear}} = \Box h_{\mu\nu} - \partial_\rho\partial_\mu h_\nu{}^\rho - \partial_\rho\partial_\nu h_\mu{}^\rho + \partial_\mu\partial_\nu h \overset{*}{=} \Box h_{\mu\nu} + \partial_\mu\partial_0 h_{\nu 0} + \partial_\nu\partial_0 h_{\mu 0} \doteq 0 ~, \quad (2.7)$$

is in turn equivalent to the three equations

$$\partial_i^2 h_{0k} + \partial_0\partial_k h_{00} \doteq 0 ~, \quad (2.8a)$$

$$(\partial_i^2 + \partial_0^2) h_{00} \doteq 0 ~, \quad (2.8b)$$

$$(\partial_k^2 - \partial_0^2) h_{ij} + \partial_0\partial_i h_{j0} + \partial_0\partial_j h_{i0} \doteq 0 ~. \quad (2.8c)$$

Note that the two terms in the l.h.s. of (2.8b) have the same sign. Consider now the integration of $h_{00}\times$ (2.8b) over the total 3D with partial integrations:

$$0 = \int d^3x \, h_{00}(\partial_i^2 + \partial_0^2) h_{00} = \int d^3x \, [-(\partial_i h_{00})^2 - (\partial_0 h_{00})^2] \quad (2.9)$$

Since the last integrand is negative definite, the only way for the integral to be zero is $h_{00} \doteq 0$. If $h_{00} \doteq 0$ is used in (2.8a), a similar reasoning leads to the unique solution $h_{0k} \doteq 0$ under the boundary condition at infinity. We thus have

$$h_{00} \doteq 0 ~, \quad h_{0i} \doteq 0 ~. \quad (2.10)$$

Using these in the remaining (2.8c), we get

$$(\partial_k^2 - \partial_0^2) h_{ij} \doteq 0 ~. \quad (2.11)$$

This means that the components $h_{ij}$ may be still propagating. However, we now recall that $h_{ij}$ can be diagonalized, such that only $h_{11}$ and $h_{22}$ remain. Now under (2.5), (2.6) and (2.10) we immediately see that

$$h_{11} \doteq -h_{22} ~, \quad \partial_1 h_{11} \doteq 0 ~, \quad \partial_2 h_{22} \doteq 0 \quad \Longrightarrow \quad \partial_i h_{11} \doteq \partial_i h_{22} \doteq 0 ~, \quad (2.12)$$

for $i = 1, 2$. This implies that there is no degree of freedom left for the whole $h_{\mu\nu}$.

(iv) The fourth method is based on the gravitino lagrangian in supergravity. Consider the gravitino kinetic term:

$$\mathcal{L}_\psi \equiv +\tfrac{1}{2}\kappa^{-2}\epsilon^{\mu\nu\rho}(\overline{\psi}_\mu D_\nu \psi_\rho) ~, \quad (2.13)$$

in $N = 1$ pure supergravity *without* matter, whose field equation yields the vanishing of the gravitino field strength $\mathcal{R}_{\mu\nu} \equiv D_\mu \psi_\nu - D_\nu \psi_\mu \doteq 0$. Due to supersymmetry, this is associated with the vanishing of the Riemann tensor.

Thus from the viewpoints of both graviton and gravitino, there seem to be no physical degree of freedom for $h_{\mu\nu}$ and $\psi_\mu$ in 3D.

In the case of (Curvature)$^2$-terms in dimensions $D \geq 4$, adding such terms do not drastically change the original physical degrees of freedom of graviton or gravitino [8][13], but they are regarded as so-called $\mathcal{O}(\alpha')$ string tension corrections [3]. From this viewpoint, it seems true in 3D that these (Curvature)$^2$-terms will not change the non-physical feature of the original Hilbert action. In fact, consider the lagrangian

$$\mathcal{L}_{R+R^2} \equiv -\tfrac{1}{4}eM^2R + \alpha e(R_{\mu\nu})^2 + \beta e R^2 \quad , \tag{2.14}$$

with $M \equiv M_{\rm Pl} \equiv \kappa^{-1}$, and real constants $\alpha$ and $\beta$. Since the Einstein tensor, and therefore, the Ricci tensor vanishes on-shell at the lowest order: $R_{\mu\nu} \doteq 0$, the newly-added Ricci tensor squared and scalar curvature-squared terms seem to vanish 'on-shell', leaving no effect on the non-physical feature of the graviton. Moreover, since the Riemann tensor is no longer independent in 3D via (2.1), we can not use the $(R_{\mu\nu\rho\sigma})^2$-term, either.

However, there is a flaw in this argument. This can be elucidated by the lagrangian of a real scalar field

$$\mathcal{L}_{\varphi^2} \equiv -\tfrac{1}{2}m^2\varphi^2 \quad . \tag{2.15}$$

With only the mass term, there is *no* physical degree of freedom. To (2.15), we add the 'higher-derivative term':

$$\mathcal{L}_{(\partial\varphi)^2} \equiv -\tfrac{1}{2}(\partial_\mu\varphi)^2 \quad . \tag{2.16}$$

After this term is added, the originally non-physical field $\varphi$ starts propagating, carrying a new physical degree of freedom. To put it differently, the original situation with (2.15) with the trivial solution $\varphi \doteq 0$ has been drastically changed by the addition of (2.16).

A similar situation is observed for a graviton in 3D. Namely, even though the graviton in the Hilbert action carries no physical degree of freedom, it starts propagating after the (Curvature)$^2$-terms are added. In other words, the original trivial situation with $R_{\mu\nu} \doteq 0$ only with $h_{\mu\nu} \doteq 0$ is changed to have propagating solution for $h_{\mu\nu}$. For example, when $\alpha = -1/4$, $\beta = +1/8$ in (2.14), the linear-order field equation of the graviton is

$$(\Box - M^2)R_{\mu\nu}\Big|_{\rm Linear} = (\Box - M^2)(\Box h_{\mu\nu} - \partial_\mu\partial_\rho h_\nu{}^\rho - \partial_\nu\partial_\rho h_\mu{}^\rho + \partial_\mu\partial_\nu h) \doteq 0 \quad . \tag{2.17}$$

Obviously, this equation has more solutions than the trivial one $R_{\mu\nu} \doteq 0$, and the graviton no longer stays just as a 'non-physical' field.

We can understand this also from the viewpoint of 'Coulomb gauge' analysis for the graviton. Our previous eq. (2.7) is entirely modified by the new factor $(\Box - M^2)$ applied from the left. For example, (2.8b) now becomes

$$(\Box - M^2)(\partial_i^2 + \partial_0^2)h_{00} \doteq 0 \quad . \tag{2.18}$$

As opposed to the previous case with no $\square$, there are now non-trivial propagating solution other than $h_{00} \doteq 0$, due to the additional factor $(\square - M^2)$.

## 3. $N=1$ Supersymmetric (Curvature)$^2$-Terms

As in higher-dimensions associated with superstring, such as in 10D [3] the possible independent (Curvature)$^2$-terms in 3D are either the (Ricci-tensor)$^2$ or (scalar curvature)$^2$-terms. In dimensions $D \geq 4$, this is because of Gauss-Bonnet theorem, dictating that (Riemann-tensor)$^2$-term can be a linear combination of (Ricci-tensor)$^2$ and (scalar curvature)$^2$-terms, up to a total divergence. In 3D, however, the Gauss-Bonnet theorem combination is not a total divergence, but instead it vanishes identically

$$(R_{\mu\nu\rho\sigma})^2 - 4(R_{\mu\nu})^2 + R^2 \equiv 0 \ , \tag{3.1}$$

because of the identity (2.1) between these curvatures. We have to keep this in mind, when considering (Curvature)$^2$-terms.

The off-shell multiplet of $N=1$ supergravity consists of the fields $(e_\mu{}^m, \psi_\mu, S)$ with $(4+4)$ degrees of freedom [4], where $e_\mu{}^m$ is the dreibein, $\psi_\mu$ is the Majorana-spinor gravitino, and $S$ is a real scalar auxiliary field. When (Curvature)$^2$-terms are added, this originally auxiliary field starts propagating with graviton and gravitino, forming two new massive multiplets with spins $\left(\frac{1}{2}, 0\right)$ under supersymmetry.

The supergravity action $I_0 \equiv \int d^3x\, \mathcal{L}_0$ has the lagrangian [4]

$$\mathcal{L}_0 \equiv -\tfrac{1}{4} M^2 e R + \tfrac{1}{2} M^2 \epsilon^{\mu\nu\rho}(\overline{\psi}_\mu D_\nu \psi_\rho) - \tfrac{1}{2} M^2 e S^2 \ , \tag{3.2}$$

where $M \equiv M_{\text{Pl}} \equiv \kappa^{-1}$, as before. Note that the last term for the $S$-field has the non-tachyonic mass.

We now consider the (Curvature)$^2$-terms. There are two possible lagrangians for supersymmetric generalizations of such (Curvature)$^2$-terms:

$$\mathcal{L}_1 \equiv -\tfrac{1}{4}\xi e (R_{\mu\nu})^2 + \tfrac{1}{8}\xi e R^2 - \tfrac{1}{4}\xi\, \epsilon^{\mu\rho\sigma}(\overline{\psi}_\mu D_\tau^2 \mathcal{R}_{\rho\sigma}) - \tfrac{1}{2}\xi e(\partial_\mu S)^2 \ , \tag{3.3a}$$

$$\mathcal{L}_2 \equiv +\tfrac{1}{32}\eta\, e R^2 - \tfrac{1}{16}\eta\, e(\overline{\mathcal{R}}_{\rho\sigma}\gamma^{\rho\sigma}\slashed{D}\gamma^{\tau\lambda}\mathcal{R}_{\tau\lambda}) - \tfrac{1}{2}\eta\, e(\partial_\mu S)^2 \ , \tag{3.3b}$$

where $\xi$ and $\eta$ are real arbitrary constants, and $\mathcal{R}_{\mu\nu} \equiv D_\mu \psi_\nu - D_\nu \psi_\mu$ is the gravitino field strength. The actions $I_1$ and $I_2$ are invariant up to trilinear-order terms under supersymmetry

$$\delta_Q e_\mu{}^m = -(\overline{\epsilon}\gamma^m \psi_\mu) \ , \tag{3.4a}$$

$$\delta_Q \psi_\mu = +D_\mu(\widehat{\omega})\epsilon + \tfrac{1}{2}(\gamma_\mu \epsilon) S \ , \tag{3.4b}$$

$$\delta_Q S = -\tfrac{1}{4}(\overline{\epsilon}\gamma^{\mu\nu}\widehat{\mathcal{R}}_{\mu\nu}) \ , \tag{3.4c}$$

where *hatted* $\widehat{\omega}$ or $\widehat{\mathcal{R}}_{\mu\nu}$ is the supercovariantization of the *unhatted* ones, as usual [13].

Some remarks are in order. Compared with conformal supergravity in 4D [8][9][13][10], our 3D system has different structures. For example, not only (3.3b), but also (3.3a) has the $S$-field kinetic term. The $\psi$-bilinear term in (3.3a) lacks the projection operator proportional to $\eta_{\mu\nu}\Box - \partial_\mu\partial_\nu$. This is because the 4D lagrangian analog [8][13][9] of our $\mathcal{L}_1$ has *no* kinetic term for the $S$-field, while both of our lagrangians $\mathcal{L}_1$ and $\mathcal{L}_2$ have the $S$-kinetic term. Another difference from the 4D case [8][13] is that the combination of Ricci tensor and scalar curvature in $\mathcal{L}_1$ in 4D is the same as that for conformal supergravity, while our 3D case is different from conformal supergravity [14][4][6].

For the investigation of propagators, we look into the bilinear terms of the total lagrangian $\mathcal{L}_{\text{tot}} \equiv \mathcal{L}_0 + \mathcal{L}_1 + \mathcal{L}_2$:

$$\mathcal{L}_{\text{tot}}\Big|_{\text{Bilinear}} = +\tfrac{1}{4}h^{\mu\nu}\left[(M^2-\xi\Box)P^{(2)}_{\mu\nu,\rho\sigma} - \{M^2-(\xi+\eta)\Box\}P^{(0,s)}_{\mu\nu,\rho\sigma}\right]\Box h^{\rho\sigma}$$
$$+\tfrac{1}{2}\overline{\psi}^\mu\left[(M^2-\xi\Box)P^{(3/2)}_{\mu\nu} - \{M^2-(\xi+\eta)\Box\}(P^{(1/2)}_{11})_{\mu\nu}\right]\not{\partial}\,\psi^\nu$$
$$-\tfrac{1}{2}S\left[M^2-(\xi+\eta)\Box\right]S \ . \tag{3.5}$$

The structures common to the gravitino and graviton/scalar show the consistency of the system. Here we construct projection operators in 3D, analogous to the corresponding ones in 4D for the graviton $h_{\mu\nu}$ [15] and for the gravitino $\psi_\mu$ [13][16], but with slightly different numerical coefficients:

$$P^{(2)}_{\mu\nu,\rho\sigma} \equiv +\tfrac{1}{2}(\theta_{\mu\rho}\theta_{\nu\sigma} + \theta_{\mu\sigma}\theta_{\nu\rho} - \theta_{\mu\nu}\theta_{\rho\sigma}) \ , \tag{3.6a}$$

$$P^{(1)}_{\mu\nu,\rho\sigma} \equiv +\tfrac{1}{2}(\theta_{\mu\rho}\omega_{\nu\sigma} + \theta_{\mu\sigma}\omega_{\nu\rho} + \theta_{\nu\rho}\omega_{\mu\sigma} + \theta_{\nu\sigma}\omega_{\mu\rho}) \ , \tag{3.6b}$$

$$P^{(0,s)}_{\mu\nu,\rho\sigma} \equiv +\tfrac{1}{2}\theta_{\mu\nu}\theta_{\rho\sigma} \ , \quad \theta_{\mu\nu} \equiv +\eta_{\mu\nu} - \Box^{-1}\partial_\mu\partial_\nu \equiv +\eta_{\mu\nu} - \omega_{\mu\nu} \ , \tag{3.6c}$$

$$P^{(0,w)}_{\mu\nu,\rho\sigma} \equiv +\omega_{\mu\nu}\omega_{\rho\sigma} \ , \quad \omega_{\mu\nu} \equiv +\Box^{-1}\partial_\mu\partial_\nu \ , \tag{3.6d}$$

$$P^{(0,sw)}_{\mu\nu,\rho\sigma} \equiv +\tfrac{1}{\sqrt{2}}\theta_{\mu\nu}\omega_{\rho\sigma} \ , \quad P^{(0,ws)}_{\mu\nu,\rho\sigma} \equiv +\tfrac{1}{\sqrt{2}}\omega_{\mu\nu}\theta_{\rho\sigma} \ , \tag{3.6e}$$

$$P^{(3/2)}_{\mu\nu} \equiv +\theta_{\mu\nu} - \tfrac{1}{2}\widehat{\gamma}_\mu\widehat{\gamma}_\nu \ , \quad \widehat{\gamma}_\mu \equiv +\gamma_\mu - \omega_\mu \ , \quad \omega_\mu \equiv +\Box^{-1}\partial_\mu\not{\partial} \ , \tag{3.6f}$$

$$(P^{(1/2)}_{11})_{\mu\nu} \equiv +\tfrac{1}{2}\widehat{\gamma}_\mu\widehat{\gamma}_\nu \ , \quad (P^{(1/2)}_{22})_{\mu\nu} \equiv +\omega_\mu\omega_\nu = +\omega_{\mu\nu} \ , \tag{3.6g}$$

$$(P^{(1/2)}_{12})_{\mu\nu} \equiv +\tfrac{1}{\sqrt{2}}\widehat{\gamma}_\mu\omega_\nu \ , \quad (P^{(1/2)}_{21})_{\mu\nu} \equiv +\tfrac{1}{\sqrt{2}}\omega_\mu\widehat{\gamma}_\nu \ , \tag{3.6h}$$

They satisfy the ortho-normality relationships

$$P^{(i,a)}_{\mu\nu,\rho\sigma}P^{(j,b)}_{\rho\sigma,\tau\lambda} = \delta^{ij}\delta^{ab}P^{(i,a)}_{\mu\nu,\tau\lambda} \ , \quad P^{(i,ab)}_{\mu\nu,\rho\sigma}P^{(j,cd)}_{\rho\sigma,\tau\lambda} = \delta^{ij}\delta^{bc}P^{(i,a)}_{\mu\nu,\tau\lambda} \ , \tag{3.7a}$$

$$P^{(i,a)}_{\mu\nu,\rho\sigma}P^{(j,bc)}_{\rho\sigma,\tau\lambda} = \delta^{ij}\delta^{ab}P^{(i,ac)}_{\mu\nu,\tau\lambda} \ , \quad P^{(i,ab)}_{\mu\nu,\rho\sigma}P^{(j,c)}_{\rho\sigma,\tau\lambda} = \delta^{ij}\delta^{bc}P^{(i,ac)}_{\mu\nu,\tau\lambda} \ , \tag{3.7b}$$

with $i,\ j = 0,\ 1,\ 2; \ a,\ b,\ c,\ d = s,\ w$, and

$$(P^{(i)}_{ab})_{\mu\nu}(P^{(j)}_{cd})_{\nu\rho} = \delta^{ij}\delta_{bc}(P^{(i)}_{ad})_{\mu\rho} \ , \tag{3.8}$$

with $i, j = 3/2, 1/2$, and $a, b, c, d = 1, 2$. These structures are parallel to the 4D cases [15][16]. The decompositions of unity are

$$\left(P^{(2)} + P^{(1)} + P^{(0,s)} + P^{(0,w)}\right)_{\mu\nu,\rho\sigma} = +\tfrac{1}{2}\eta_{\mu\rho}\eta_{\nu\sigma} + \tfrac{1}{2}\eta_{\mu\sigma}\eta_{\nu\rho} \ , \tag{3.9a}$$

$$\left(P^{(3/2)} + P^{(1/2)}_{11} + P^{(1/2)}_{22}\right)_{\mu\nu} = +\eta_{\mu\nu} \ . \tag{3.9b}$$

Relevantly, some useful relationships for the bilinear kinetic terms are

$$eR\Big|_{\text{Bilinear}} = -h^{\mu\nu}(P^{(2)} - P^{(0,s)})_{\mu\nu,\rho\sigma}\Box h^{\rho\sigma} + \text{(total divergence)} \ , \tag{3.10a}$$

$$(R_{\mu\nu})^2 - \tfrac{1}{2}R^2 = +h^{\mu\nu}(P^{(2)} - P^{(0,s)})_{\mu\nu,\rho\sigma}\Box^2 h^{\rho\sigma} \ , \tag{3.10b}$$

$$R^2 = +8h^{\mu\nu}P^{(0,s)}_{\mu\nu,\rho\sigma}\Box^2 h^{\rho\sigma} \ , \tag{3.10c}$$

$$(R_{\mu\nu})^2 = +h^{\mu\nu}(P^{(2)} + 3P^{(0,s)})_{\mu\nu,\rho\sigma}\Box^2 h^{\rho\sigma} \ , \tag{3.10d}$$

$$\epsilon^{\mu\rho\sigma}(\overline{\psi}_\mu \mathcal{R}_{\rho\sigma}) = +2[\overline{\psi}^\mu(P^{(3/2)} - P^{(1/2)}_{11})_{\mu\nu}\slashed{\partial}\psi^\nu] \ , \tag{3.10e}$$

$$\epsilon^{\mu\rho\sigma}(\overline{\psi}_\mu \Box \mathcal{R}_{\rho\sigma}) = +2[\overline{\psi}^\mu(P^{(3/2)} - P^{(1/2)}_{11})_{\mu\nu}\Box\slashed{\partial}\psi^\nu] \ , \tag{3.10f}$$

$$(\overline{\mathcal{R}}_{\mu\nu}\gamma^{\mu\nu}\slashed{\partial}\gamma^{\rho\sigma}\mathcal{R}_{\rho\sigma}) = -8[\overline{\psi}^\mu(P^{(1/2)}_{11})_{\mu\nu}\Box\slashed{\partial}\psi^\nu] \ . \tag{3.10g}$$

These expressions are valid up to trilinear-order terms, and total divergences. Note also that these are 3D analogs of the corresponding ones in 4D [15][16].

The propagators for $h_{\mu\nu}$, $\psi_\mu$ and $S$-fields can be obtained by inverting the spin blocks in the total lagrangian (3.5), following [13][16]

$$\langle Th_{\mu\nu}h_{\rho\sigma}\rangle = +\frac{P^{(2)}_{\mu\nu,\rho\sigma} - P^{(0,s)}_{\mu\nu,\rho\sigma}}{\Box} - \frac{P^{(2)}_{\mu\nu,\rho\sigma}}{\Box - \xi^{-1}M^2} + \frac{P^{(0,s)}_{\mu\nu,\rho\sigma}}{\Box - \frac{M^2}{\xi+\eta}} \ , \tag{3.11a}$$

$$\langle T\psi_\mu\psi_\nu\rangle = +\frac{P^{(3/2)}_{\mu\nu} - (P^{(1/2)}_{11})_{\mu\nu}}{\slashed{\partial}} - \frac{P^{(3/2)}_{\mu\nu}}{2(\slashed{\partial} - \xi^{-1/2}M)} - \frac{P^{(3/2)}_{\mu\nu}}{2(\slashed{\partial} + \xi^{-1/2}M)}$$
$$+ \frac{(P^{(1/2)}_{11})_{\mu\nu}}{2\left(\slashed{\partial} - \frac{M}{\sqrt{\xi+\eta}}\right)} + \frac{(P^{(1/2)}_{11})_{\mu\nu}}{2\left(\slashed{\partial} + \frac{M}{\sqrt{\xi+\eta}}\right)} \ , \tag{3.11b}$$

$$\langle TSS\rangle = +\frac{1}{\Box - \frac{M^2}{\xi+\eta}} \ . \tag{3.11c}$$

Even though we omitted the inessential factors, such as $1/4$, we maintain the right signs for these propagators, in order to see negative energy ghosts. The common mass poles at $\mathcal{M} = M/\sqrt{\xi}$ or $\mathcal{M} = M/\sqrt{\xi+\eta}$ for different fields support the validity of this result.

Note that the massless poles with $P^{(2)}$, $P^{(0,s)}$, $P^{(3/2)}$ and $P^{(1/2)}_{11}$ correspond to the original massless supergravity multiplet with a graviton and a gravitino, similarly to the 4D case [8][13]. In particular, the combinations $P^{(2)} - P^{(0,s)}$ and $P^{(3/2)} - P^{(1/2)}_{11}$ are parallel to the corresponding terms in 4D [8][13]. The negative signs for the massless poles with $P^{(0,s)}$ and $P^{(1/2)}$ do not pose any problem, because they are parts of the supergravity multiplet, just as in the 4D case [8][13]. The overall positive sign with the relative sign between $\Box$ and $M^2/(\xi+\eta)$ for the $S$-propagator correspond to the positive energy with non-tachyonic masses, which can be the 'reference sign' for the $h_{\mu\nu}$ propagators. For the overall sign for the $\psi$-propagators, the positive sign corresponds to the positive energy.

There are massive poles for the graviton propagator with the (mass)$^2$, i.e., $\mathcal{M}^2 = M^2/\xi$ and $M^2/(\xi+\eta)$. The same pattern is also found for the gravitino propagator with the $\mathcal{M} = \pm M/\sqrt{\xi}$ and $\pm M/\sqrt{\xi+\eta}$. These masses are related to each other under supersymmetry. We also see that the propagator signs for $\mathcal{M}^2 = M^2/\xi$ or $\mathcal{M} = \pm M/\sqrt{\xi}$ have negative energy. In order to exclude these negative energy propagators, we have to impose the condition $\xi = 0$, so that these poles will disappear with infinitely heavy masses. In other words, only the lagrangian $\mathcal{L}_2$ is acceptable without negative energy ghosts. In this case, since the $S$-kinetic term gets $\eta$ in front, we can normalize $\eta = +1$. After all, we have

$$\xi = 0 \ , \quad \eta = +1 \ . \tag{3.12}$$

In this case, all the propagators are simplified, and there is no negative energy ghost among the massive propagators:

$$\langle Th_{\mu\nu}h_{\rho\sigma}\rangle = \ + \frac{P^{(2)}_{\mu\nu,\rho\sigma} - P^{(0,s)}_{\mu\nu,\rho\sigma}}{\Box} + \frac{P^{(0,s)}_{\mu\nu,\rho\sigma}}{\Box - M^2} \ , \tag{3.13a}$$

$$\langle T\psi_\mu\psi_\nu\rangle = \ + \frac{P^{(3/2)}_{\mu\nu} - (P^{(1/2)}_{11})_{\mu\nu}}{\slashed{\partial}} + \frac{(P^{(1/2)}_{11})_{\mu\nu}}{2(\slashed{\partial} - M)} + \frac{(P^{(1/2)}_{11})_{\mu\nu}}{2(\slashed{\partial} + M)} \ , \tag{3.13b}$$

$$\langle TSS\rangle = \ + \frac{1}{\Box - M^2} \ . \tag{3.13c}$$

Now all the propagating components are physical, forming the massless supermultiplet of spins $\left(2, \frac{3}{2}\right)$ by $h_{\mu\nu}$ and $\psi_\mu$, and two massive supermultiplets with spins $\left(\frac{1}{2}, 0\right)$ with the mass $\mathcal{M} = M$. The first of these is a spin 0 from $h_{\mu\nu}$ and a spin 1/2 from $\psi_\mu$, while the second is from a spin 1/2 from $\psi_\mu$ and one spin 0 from $S$. These components form $2 \times (1+1)$ degrees of freedom.

Some readers may wonder, if the two signs for the mass $\mathcal{M} = \pm M$ for the spin 1/2 propagator cause any problem with the positive definiteness of energy. In 4D, for a Majorana or Dirac spinor, the signature of the mass term does not matter, because we can always perform the replacement $\psi \to i\gamma_5\psi$, leaving the kinetic term intact, while flipping the sign of the mass term. In 3D, despite the absence of the analog of the $\gamma_5$-matrix, the mass-term sign does not pose any problem. There are two independent ways to understand this. The first way is to consider the 'dynamical' energy-momentum tensor for the kinetic and mass terms for a spin $1/2$ Majorana field $\chi$:

$$\mathcal{L}_\chi \equiv +\tfrac{1}{2} e e_m{}^\mu (\overline{\chi}\gamma^m D_\mu \chi) + \tfrac{1}{2} m e (\overline{\chi}\chi) \ , \tag{3.14}$$

where $D_\mu$ contains the usual Lorentz connection $\omega_\mu{}^{rs}(e)$ in terms of the dreibein $e_\mu{}^m$. The dynamical energy-momentum tensor is obtained by varying the linearized metric:

$$T_{\mu\nu} \equiv \frac{\delta \mathcal{L}_\chi}{\delta h^{\mu\nu}} = -\tfrac{1}{2}\eta_{\mu\nu}\overline{\chi}(\slashed{D}\chi + m\chi) + \tfrac{1}{2}(\overline{\chi}\gamma_{(\mu}D_{\nu)}\chi) \ . \tag{3.15}$$

The point is that the first term in (3.15) vanishes upon the $\chi$-field equation, independent of the signature of $m$. Therefore, the difference between $m > 0$ and $m < 0$ does not affect the positive definiteness of the $T^{00}$-component.

The second way is more intuitive, based on the $N = 1$ scalar multiplet $(\chi, \varphi)$ with the action

$$I_{\chi,\varphi} \equiv \int d^3x \left[ -\tfrac{1}{2}(\partial_\mu \varphi)^2 + \tfrac{1}{2}(\overline{\chi}\slashed{\partial}\chi) - \tfrac{1}{2}m^2 \varphi^2 + \tfrac{1}{2}m(\overline{\chi}\chi) \right] \ , \tag{3.16}$$

invariant under $N = 1$ global supersymmetry

$$\delta_Q \varphi = +\tfrac{1}{\sqrt{2}}(\overline{\epsilon}\chi) \ , \qquad \delta_Q \chi = -\tfrac{1}{\sqrt{2}}(\gamma^\mu \epsilon)\partial_\mu \varphi + \tfrac{1}{\sqrt{2}} m\, \epsilon\, \varphi \ . \tag{3.17}$$

The validity of supersymmetric invariance $\delta_Q I_{\chi,\varphi} = 0$ is independent of the sign of $m$. Since the scalar $\varphi$ has the positive definite energy with a non-tachyonic mass, there is no problem with its super-partner $\chi$ for both cases of $m > 0$ and $m < 0$, as guaranteed by supersymmetry.

## 4. $N = 2$ Supersymmetric (Curvature)$^2$-Terms

Once we have established $N = 1$ supersymmetric (Curvature)$^2$-terms, it is straightforward to generalize it to $N = 2$ supergravity. The off-shell $N = 2$ supergravity multiplet consists of $(e_\mu{}^m, \psi_\mu, \psi_\mu^*, A_\mu, B, B^*)$ with $(8+8)$ degrees of freedom [5][6], where the gravitino is now a Dirac spinor $\psi_\mu \equiv \psi_\mu^{(1)} + i\,\psi_\mu^{(2)}$ in terms of two Majorana spinors $\psi_\mu^{(1)}$ and $\psi_\mu^{(2)}$, so we have to distinguish the *starred* $\psi_\mu^* = \psi_\mu^{(1)} - i\,\psi_\mu^{(2)}$ from the *unstarred* $\psi_\mu \equiv \psi_\mu^{(1)} + i\,\psi_\mu^{(2)}$.[6] The auxiliary fields are the real vector $A_\mu$ and the complex scalar $B$ with its complex conjugate $B^*$. These auxiliary fields resemble those in $N = 1$ supergravity in 4D [7][13][10], because $N = 1$ supergravity in 3D is directly obtained from the latter by a simple dimensional reduction.

As in the $N = 1$ case, we consider the total action $I_{\text{tot}} \equiv I_0 + I_1 + I_2$ in terms of three actions $I_0$, $I_1$ and $I_2$, where the corresponding lagrangians are

$$\mathcal{L}_0 \equiv -\tfrac{1}{4}M^2 eR + \tfrac{1}{2}M^2 \epsilon^{\mu\nu\rho}\left[(\overline{\psi}_\mu^* D_\nu \psi_\rho) + (\overline{\psi}_\mu D_\nu \psi_\rho^*)\right] + \tfrac{1}{2}M^2 e A_\mu^2 - \tfrac{1}{2}M^2 e|\partial_\mu B|^2 \ , \quad (4.1a)$$

---

[6] See eq. (4.3) for practical examples.

$$\mathcal{L}_1 \equiv -\tfrac{1}{4}\xi e(R_{\mu\nu})^2 + \tfrac{1}{8}\xi e R^2 - \tfrac{1}{4}\xi\,\epsilon^{\mu\rho\sigma}\left[(\overline{\psi}^*_\mu D^2_\tau \mathcal{R}_{\rho\sigma}) + (\overline{\psi}_\mu D^2_\tau \mathcal{R}^*_{\rho\sigma})\right]$$
$$+ \tfrac{1}{2}\xi e(D_\mu A_\nu)^2 - \tfrac{1}{2}\xi e|\partial_\mu B|^2 \ , \tag{4.1b}$$

$$\mathcal{L}_2 \equiv +\tfrac{1}{32}\eta\,e R^2 - \tfrac{1}{16}\eta\,e\left[(\overline{\mathcal{R}}^*_{\rho\sigma}\gamma^{\rho\sigma}\slashed{D}\gamma^{\tau\lambda}\mathcal{R}_{\tau\lambda}) + (\overline{\mathcal{R}}_{\rho\sigma}\gamma^{\rho\sigma}\slashed{D}\gamma^{\tau\lambda}\mathcal{R}^*_{\tau\lambda})\right]$$
$$+ \tfrac{1}{2}\eta\,e(D_\mu A^\mu)^2 - \tfrac{1}{2}\eta\,e|\partial_\mu B|^2 \ . \tag{4.1c}$$

The actions $I_0$, $I_1$ and $I_2$ are invariant up to trilinear terms under supersymmetry [5][6]

$$\delta_Q e_\mu{}^m = -(\overline{\epsilon}^*\gamma^m\psi_\mu) - (\overline{\epsilon}\gamma^m\psi^*_\mu) \ , \tag{4.2a}$$

$$\delta_Q \psi_\mu = +D_\mu(\widehat{\omega})\epsilon + \tfrac{i}{2}(\gamma^\nu\gamma_\mu\epsilon)A_\nu + \tfrac{i}{2}(\gamma_\mu\epsilon)B \ , \tag{4.2b}$$

$$\delta_Q \psi^*_\mu = +D_\mu(\widehat{\omega})\epsilon^* - \tfrac{i}{2}(\gamma^\nu\gamma_\mu\epsilon^*)A_\nu - \tfrac{i}{2}(\gamma_\mu\epsilon^*)B^* \ , \tag{4.2c}$$

$$\delta_Q A_\mu = +\tfrac{i}{4}(\overline{\epsilon}^*\gamma^{\rho\sigma}\gamma_\mu\widehat{\mathcal{R}}_{\rho\sigma}) - \tfrac{i}{4}(\overline{\epsilon}\gamma^{\rho\sigma}\gamma_\mu\widehat{\mathcal{R}}^*_{\rho\sigma}) \ , \tag{4.2d}$$

$$\delta_Q B = +\tfrac{i}{2}(\overline{\epsilon}\gamma^{\mu\nu}\widehat{\mathcal{R}}_{\mu\nu}) \ , \quad \delta_Q B^* = -\tfrac{i}{2}(\overline{\epsilon}^*\gamma^{\mu\nu}\widehat{\mathcal{R}}^*_{\mu\nu}) \ . \tag{4.2e}$$

Due to the Dirac nature of the spinors, we need a special care for the *star*-symbols, which are different from those used for Majorana bilinears. For example, the second kinetic term of the gravitino is just the complex conjugate of the first one. Typical examples are such as

$$(\overline{\epsilon}^*\gamma^m\psi_\mu) = (\overline{\epsilon}^{(1)} - i\,\overline{\epsilon}^{(2)})\gamma^m(\psi_\mu^{(1)} + i\,\psi_\mu^{(2)}) \ ,$$
$$(\overline{\epsilon}^*\gamma^m\psi_\mu)^* = (\overline{\epsilon}^{(1)} + i\,\overline{\epsilon}^{(2)})\gamma^m(\psi_\mu^{(1)} - i\,\psi_\mu^{(2)}) = (\overline{\epsilon}\gamma^m\psi^*_\mu) \ , \tag{4.3}$$

where the Dirac spinors $\epsilon$ and $\psi_\mu$ are expressed in terms of the Majorana spinors $\epsilon^{(1)}$, $\epsilon^{(2)}$, $\psi_\mu^{(1)}$ and $\psi_\mu^{(2)}$.

As in the $N=1$ case, all the bilinear terms in $\mathcal{L}_{\text{tot}}$ can be re-expressed in terms of projection operators. The only subtlety is the $A_\mu$-bilinear term rearranged as

$$(A\text{-Bilinear Terms}) = +\tfrac{1}{2}A^\mu\left[(M^2 - \xi\Box)P^{(\text{T})}_{\mu\nu} + \left\{M^2 - (\xi+\eta)\Box\right\}P^{(\text{L})}_{\mu\nu}\right]A^\nu \ , \tag{4.4}$$

where $P^{(\text{T})}_{\mu\nu} \equiv \theta_{\mu\nu}$ and $P^{(\text{L})}_{\mu\nu} \equiv \omega_{\mu\nu}$.

These bilinear terms can be inverted to yield the propagators

$$\langle T h_{\mu\nu} h_{\rho\sigma}\rangle = +\frac{P^{(2)}_{\mu\nu,\rho\sigma} - P^{(0,s)}_{\mu\nu,\rho\sigma}}{\Box} - \frac{P^{(2)}_{\mu\nu,\rho\sigma}}{\Box - \xi^{-1}M^2} + \frac{P^{(0,s)}_{\mu\nu,\rho\sigma}}{\Box - \frac{M^2}{\xi+\eta}} \ , \tag{4.5a}$$

$$\langle T\psi_\mu\overline{\psi}^*_\nu\rangle = +\frac{P^{(3/2)}_{\mu\nu} - (P^{(1/2)}_{11})_{\mu\nu}}{\slashed{\partial}} - \frac{P^{(3/2)}_{\mu\nu}}{2(\slashed{\partial} - \xi^{-1/2}M)} - \frac{P^{(3/2)}_{\mu\nu}}{2(\slashed{\partial} + \xi^{-1/2}M)}$$
$$+ \frac{(P^{(1/2)}_{11})_{\mu\nu}}{2\left(\slashed{\partial} - \frac{M}{\sqrt{\xi+\eta}}\right)} + \frac{(P^{(1/2)}_{11})_{\mu\nu}}{2\left(\slashed{\partial} + \frac{M}{\sqrt{\xi+\eta}}\right)} = \langle T\psi^*_\mu\overline{\psi}_\nu\rangle \ , \tag{4.5b}$$

$$\langle TA_\mu A_\nu \rangle = \ -\frac{P^{(T)}_{\mu\nu}}{\Box - \xi^{-1}M^2} - \frac{P^{(L)}_{\mu\nu}}{\Box - \frac{M^2}{\xi+\eta}} \quad, \tag{4.5c}$$

$$\langle TBB^* \rangle = \ +\frac{1}{\Box - \frac{M^2}{\xi+\eta}} = \langle TB^*B \rangle \ . \tag{4.5d}$$

These are up to inessential positive overall constants, as in the previous $N=1$ case.

As in the $N=1$ case, we can get rid of the negative energy ghosts with the poles at $\mathcal{M} = M^2/\xi$ or $\mathcal{M} = M/\sqrt{\xi}$, together with the normalization of the $BB^*$-propagator, as

$$\xi = 0 \ , \quad \eta = +1 \ . \tag{4.6}$$

In such a case, the propagators are simplified as

$$\langle Th_{\mu\nu}h_{\rho\sigma} \rangle = \ +\frac{P^{(2)}_{\mu\nu,\rho\sigma} - P^{(0,s)}_{\mu\nu,\rho\sigma}}{\Box} + \frac{P^{(0,s)}_{\mu\nu,\rho\sigma}}{\Box - M^2} \quad, \tag{4.7a}$$

$$\langle T\psi^*_\mu \overline{\psi}_\nu \rangle = \ +\frac{P^{(3/2)}_{\mu\nu} - (P^{(1/2)}_{11})_{\mu\nu}}{\slashed{\partial}} + \frac{(P^{(1/2)}_{11})_{\mu\nu}}{2(\slashed{\partial}-M)} + \frac{(P^{(1/2)}_{11})_{\mu\nu}}{2(\slashed{\partial}+M)} = \langle T\psi_\mu \overline{\psi}^*_\nu \rangle \ , \tag{4.7b}$$

$$\langle T(\partial_\mu A^\mu)(\partial_\nu A^\nu) \rangle = \ +\frac{1}{\Box - M^2} \quad, \tag{4.7c}$$

$$\langle TBB^* \rangle = \ +\frac{1}{\Box - M^2} = \langle TB^*B \rangle \ . \tag{4.7d}$$

Even though the overall sign for the $A_\mu A_\nu$-propagator in (4.5c) is negative, we can interpret that the longitudinal component $\partial_\mu A^\mu$ has positive definite propagator as in (4.7c), after a partial integration at the bilinear lagrangian level.

As in the $N=1$ case, all the propagating components are physical, forming $N=2$ supermultiplets. All the massless components form the massless $N=2$ supergravity multiplet $\left(2, \frac{3}{2}, \frac{3}{2}\right)$. From the Dirac spinors $\psi_\mu$ and $\overline{\psi}_\mu$, there are in total four spin $1/2$ components with $\mathcal{M} = M$, while $B$ and $B^*$ contribute two spin $0$ components with $\mathcal{M}^2 = M^2$, while $\partial_\mu A^\mu$ counts as one spin $0$ with $\mathcal{M}^2 = M^2$. Another spin $0$ component with $\mathcal{M}^2 = M^2$ comes from $h_{\mu\nu}$. Eventually, these form two massive $N=2$ multiplets of spins $\left(\frac{1}{2}, \frac{1}{2}, 0, 0\right)$ of the mass $M$ with $2 \times (2+2)$ degrees of freedom. To be more specific, the two components of spin $1/2$ in $\psi_\mu$ and the complex field $B$ form the first $N=2$ multiplet $\left(\frac{1}{2}, \frac{1}{2}, 0, 0\right)$, while the spin $0$ component in $h_{\mu\nu}$, the remaining two components with spin $1/2$ in $\psi_\mu$, and $\partial_\mu A^\mu$ form the second $N=2$ multiplet $\left(\frac{1}{2}, \frac{1}{2}, 0, 0\right)$. Compared with our previous $N=1$ case in 3D, the total degrees of freedom are doubled, because of the new additional 'auxiliary' field components $\mathcal{I}m B$ and $\partial_\mu A^\mu$ together with $i(\psi_\mu - \psi^*_\mu)$.

There are differences as well as similarities compared with the $N=1$ supersymmetrization of (Curvature)$^2$-terms in 4D [8][13][9]. One similarity is, of course, essentially the same off-shell field content, *i.e.*, our complex field $B$ is equivalent to two scalars $S$ and $P$ used in $N=1$ supergravity in 4D [7][13][10]. This is reflected in the pattern of our two $N=2$ multiplets $\left(\frac{1}{2}, \frac{1}{2}, 0, 0\right)$ formed by the spin $1/2$ and spin $0$ contents out of the fields $h_{\mu\nu}$, $B$, $\psi_\mu$ and $\partial_\mu A^\mu$. Another similarity is that the longitudinal mode $\partial_\mu A^\mu$ is propagating in the total action $I_0 + I_2$. The negative energy ghosts can be avoided, by avoiding the action $I_1$ both in 3D and 4D [8]. The difference is, of course, that the graviton and gravitino in 3D are not physical without $I_1$ or $I_2$, but start propagating only in the presence of $I_1$ or $I_2$.

## 5. Summary and Concluding Remarks

In this paper, we have investigated the effect of (Curvature)$^2$-terms on $N=1$ and $N=2$ supergravity in 3D. Interestingly, we have found that only the (Scalar Curvature)$^2$-term can be supersymmetrized both in $N=1$ and $N=2$, without negative energy ghost poles.

We have first presented two supersymmetric lagrangians for (Curvature)$^2$-terms for $N=1$ supergravity in 3D. Due to the relationship (2.1) among curvature tensors, there are only two possible lagrangians (3.3a) and (3.3b). Subsequently, we have expressed the bilinear-order terms in $\mathcal{L}_{\rm tot} = \mathcal{L}_0 + \mathcal{L}_1 + \mathcal{L}_2$ in terms of projection operators as in (3.5). Based on this, we have obtained the propagators for the graviton $h_{\mu\nu}$, gravitino $\psi_\mu$ and scalar field $S$, as in (3.11). In order to avoid negative energy ghosts, while maintaining the canonical kinetic term for $S$, we have to impose the condition $\xi = 0$, $\eta = +1$. In such a case, the propagators are drastically simplified as in (3.13). In this final form, we see that the spin $0$ part of $h_{\mu\nu}$, spin $1/2$ part of $\psi_\mu$ and the spin $0$ field $S$ form two massive $N=1$ multiplets of spins $\left(\frac{1}{2}, 0\right)$ with $2 \times (1+1)$ degrees of freedom, consistent with supersymmetry.

A similar analysis has been applied to the $N=2$ case with the supersymmetric lagrangians (4.1), yielding the propagators (4.5). We found again the condition $\xi = 0$, $\eta = +1$ in order to avoid negative energy ghosts as in (4.6). The resulting propagating physical components are doubled compared with the $N=1$ case, namely, we have two massive $N=2$ supermultiplets of spins $\left(\frac{1}{2}, \frac{1}{2}, 0, 0\right)$ with $2 \times (2+2)$ degrees of freedom. The new contributions are from $i(\psi_\mu - \psi_\mu^*)$, $\mathcal{I}m\, B$ and $\partial_\mu A^\mu$.

Our result here may shed some light on the problem of (Curvature)$^2$-terms in 11D supergravity [17]. This is not only due to the similarity between 11D and 3D for fermionic structures, but also because of 3D serving as the world-volume for supermembrane theory [18]. From these viewpoints, our results offer new revenues for investigations in extended supergravity in 3D.

This work is supported in part by NSF Grant # 0308246.